\documentclass[pra,twocolumn,showpacs,superscriptaddress,aps]{revtex4}

\usepackage{amsfonts}
\usepackage{amsmath}
\usepackage{amssymb}
\usepackage{graphicx}
\usepackage{dcolumn}
\usepackage{dsfont}
\usepackage{times}
\usepackage{color}

\begin{document}

\title{Beating dark-dark solitons and Zitterbewegung in spin-orbit coupled Bose-Einstein condensates }

\author{V. Achilleos}
\affiliation{Department of Physics, University of Athens, Panepistimiopolis,
Zografos, Athens 15784, Greece}
\author{D. J. Frantzeskakis}
\affiliation{Department of Physics, University of Athens, Panepistimiopolis,
Zografos, Athens 15784, Greece}
\author{P. G. Kevrekidis}
\affiliation{Department of Mathematics and Statistics, University of Massachusetts,
Amherst MA 01003-4515, USA}

\begin{abstract}
We present families of beating dark-dark solitons in 
spin-orbit (SO) coupled Bose-Einstein condensates. 
These families consist of solitons 
residing simultaneously in the two bands of the energy spectrum. 
The soliton components are characterized by two different spatial and temporal scales, which are 
identified by a multiscale expansion method. The solitons are ``beating'' ones, as they perform 
density oscillations with a characteristic frequency, relevant to Zitterbewegung (ZB). 
We find that spin oscillations may occur, depending on the parity of each soliton branch, which consequently 
lead to ZB oscillations of the beating dark solitons.
Analytical results are corroborated by numerical simulations, illustrating the robustness of the solitons. 
\end{abstract}

\pacs{05.45.Yv, 03.75.Lm, 03.75.Mn}

\maketitle

\section{Introduction}
Solitons in multi-component systems is a fascinating topic with a rich history 
spanning diverse areas, including Bose-Einstein condensates (BECs) in atomic
physics~\cite{BECBOOK,nonlin}, optical fibers and photonic crystals
in nonlinear optics~\cite{yuri}, and integrable systems in mathematical
physics~\cite{ablowitz}. Different families of ``vector solitons'' have been predicted 
in these settings, and also observed in experiments. 
In the context of atomic BECs, the first relevant 
experimental observation refers to the so-called dark-bright (DB) solitons, 
which were created in a repulsive $^{87}$Rb BEC binary mixture, using 
a phase-imprinting method~\cite{hamburg} or in two counter-flowing $^{87}$Rb BECs \cite{pe}. 
These structures are composed by a dark soliton (density dip) in the first component, 
which creates --through the nonlinear coupling-- an effective trapping potential that localizes a
bright soliton (density bump) in the second component. 
Other types of vector solitons, namely ``beating'' dark-dark (DD) solitons, 
were also experimentally observed \cite{engels3}. These solitons have the form of 
two nonlinearly coupled dark solitons, that perform
a breathing oscillation between their densities~\cite{engels3,DD}. Beating DD solitons are closely
related to the DB solitons, as they emerge via a rotation of a DB soliton
in field configuration space, due to the internal symmetry of the model \cite{park} 
(which, in the homogeneous setting, is actually the so-called Manakov system \cite{manakov}).

The recent experimental realization of spin-orbit coupled (SOC) BECs~\cite{spiel2,spiel1} 
in a multicomponent $^{87}$Rb condensate has motivated further studies on vector solitons and other nonlinear 
waves in this setting. First, we should note that SOC-BECs are of particular interest mainly due 
to their relevance in systems having a Dirac-like energy band structure.
Such systems include, among others, trapped ions~\cite{ions}, 
photonic crystals~\cite{photcr}, sonic crystals~\cite{socr}, graphene~\cite{graphene}, and 
optical waveguides~\cite{optics}.
It is, therefore, not surprising that 
many studies have been devoted to nonlinear structures that may emerge in SOC-BECs. Indeed, 
structures with nontrivial topological properties, such as vortices~\cite{xuhan,spielpra}, 
Skyrmions \cite{skyrm}, Dirac monopoles \cite{dirac} and dark solitons~\cite{brand,eplva}, 
as well as self-trapped states~\cite{santos}, bright solitons~\cite{prlva} and also gap-solitons~\cite{gapkono}, 
were predicted to occur in SOC-BECs. 

Many of the interesting phenomena that can occur in SOC-BECs are due to their unique energy spectrum, 
which is composed of two branches (in the two-component system). The structure of their spectrum 
is exploited in fundamental effects, such as the Zitterbewegung (ZB) oscillations~\cite{sch}, 
Klein tunneling, or even pair-production~\cite{greiner}. Regarding ZB oscillations, it should be mentioned that 
they appear when energy eigenstates from different bands coexist and oscillations of 
the mean velocity occur. This phenomenon was observed in recent SOC-BEC experiments \cite{zbbec2}, and illustrates
the versatility of atomic condensates in simulating phenomena of the relativistic Dirac equation.

Here, we exploit both the inherent nonlinearity of BECs and the nature of the spin-orbit energy spectrum, 
and present beating DD soliton families in SOC-BECs. We show that these solitons 
emerge from states that coexist in the pertinent upper- and lower-energy band of the linear spectrum; 
the solitons are ``beating'', as they perform density oscillations with a characteristic 
frequency (the ZB frequency \cite{sch,zbbec2}). The beating DD solitons that we present in this work 
are structurally different from the ones studied in Refs.~\cite{engels3,DD}, since they exist  
only due to the particular band structure of SOC-BECs (and not due to the internal symmetry of 
the model, as mentioned above). 
Using a multiscale expansion method, we show that these 
solitons can be systematically constructed via DB solitons satisfying a system of two coupled 
nonlinear Schr\"{o}dinger (NLS) equations, governing the evolution of states in each of 
the two energy bands. The derived solitons feature two different spatial length scales 
(again stemming from the structure of the energy spectrum), as well as two different temporal scales, 
corresponding to the fast ZB oscillations and the long-time dynamics of the soliton centers. 
A connection between the beating DD-solitons and ZB oscillations observed in experiments
is also provided. We find that soliton branches of 
definite parity in the two solution branches exhibit 
spin oscillations, and also 
ZB oscillations, while branches of non-definite parity have a vanishing
 z-spin component. 

The paper is organized as follows. In Section II we present the model and our analytical 
approximations, namely the derivation of the effective NLS system. Section III is devoted to the 
analytical and numerical study of the soliton solutions, while in Section IV we discuss the ZB 
oscillations. Finally, in Section V we summarize our conclusions.

\section{Model and analytical considerations}
We consider a quasi one-dimensional SO coupled BEC
for which, in the mean-field approximation, the equation of motion is expressed in dimensionless form as \cite{spiel1,zbbec2}:
\begin{eqnarray}
i\partial_t\mathbf{\Psi}(x,t)=\mathcal{H}\mathbf{\Psi}(x,t), \quad
\mathcal{H}=\mathcal{H}_1+\mathcal{H}_{\rm int},
\label{ham}
\end{eqnarray}
where $\mathbf{\Psi}(x,t)\equiv [\psi_\uparrow(x), \psi_\downarrow(x)]^T$ and  
$\psi_{\uparrow,\downarrow}$ denote the wavefunctions of the two pseudo-spin components, 
normalized to the respective number of atoms $N_{\downarrow,\uparrow}=\int |\psi_{\downarrow,\uparrow}|^2dx$.
The non-interacting part of the Hamiltonian $\mathcal{H}_1$ is given by 
\begin{eqnarray}
\!\!\!\!\!\mathcal{\mathcal{H}}_1\!\!\!&=&\!\frac{1}{2}(\hat{p}_x \mathds{1}-k_L{\sigma}_z)^2+V_{\rm tr}(x) \mathds{1}+\Omega{\sigma}_x+\delta{\sigma}_z,
\label{ham0}
\end{eqnarray}
where length is measured in units of $a_\perp=\sqrt{\hbar/(m
\omega_\perp)}$ ($m$ being the atomic mass), energy in units of $\hbar \omega_\perp$, and time in units of 
$\omega_\perp^{-1}$. Additionally, $\hat{\sigma}_{x,z}$ are 
the usual $2\times2$ Pauli matrices,
$\mathds{1}$ is the unit matrix, $k_L=(2\pi/\lambda_L)a_\perp$, where $\lambda_L$ is the wavelength of the Raman 
laser coupling the two hyperfine states, $\Omega=\sqrt{2}\Omega_R/\hbar\omega_\perp$ 
where $\Omega_R$ is the Rabi frequency, $\delta$ is the energy offset due to Raman detuning, 
and $V_{\rm tr}(x)= (\omega_x/\omega_\perp)^2 x^2/2$ is the parabolic trap.
On the other hand, the interaction Hamiltonian $\mathcal{H}_{\rm int}$ reads:
\begin{eqnarray}
\mathcal{{H}}_{\rm int}\!&=&\!\left [ \begin{array}{c c}
|\psi_\uparrow|^2 +\beta |\psi_\downarrow|^2 & 0 \\
0 & \beta|\psi_\uparrow|^2 + |\psi_\downarrow|^2
\end{array}
\right], 
\label{Alpha}
	\end{eqnarray}
where $\beta=g_{11}/g_{12}$, with coupling constants $g_{ij}=\alpha_{ij}/\alpha_{11}$ ($i,j=1,2$) 
defined by the $s$-wave scattering lengths $\alpha_{ij}$; the latter are assumed to be positive, accounting for repulsive interactions 
(densities are measured in units of $2\alpha_{11}$). 

%
%

To find approximate solutions of Eq.~(\ref{ham}), we introduce the following asymptotic expansion:
\begin{eqnarray}
\mathbf{\Psi}=e^{ikx}\sum_{n=1}^{\infty}\epsilon^n\left(\mathbf{u}^{(n)}_+\phi^{(n)}_++\mathbf{u}^{(n)}_-\phi^{(n)}_-\right), 
\label{expansion}
\end{eqnarray}
where $\mathbf{u}^{(n)}_{\pm}(t)$ are time-dependent unknown vectors, $\phi^{(n)}_{+}(X_+,T)$ and $\phi^{(n)}_{-}(X_-,T)$ 
are unknown scalars depending on the slow scale variables $X_{\pm}=\epsilon(x-v_{g}^{\pm}t)$ and $T=\epsilon^2 t$,  while 
$\epsilon$ is a formal small parameter. Accordingly, 
the Hamiltonian~(\ref{ham}) is expanded as  
$\mathcal{H}=\sum_{n=1}^{\infty}\epsilon^n \mathcal{H}_n$, where the first three terms of the expansion are:
%
\begin{eqnarray}
\mathcal{H}_1&=&-\frac{1}{2}\partial_x^2-(ik_L\partial_x-\delta)\sigma_z+\Omega\sigma_x, 
\label{h1}\\
\mathcal{H}_2&=&-\partial_x(\partial_{X_1}+\partial_{X_2})-ik_L(\partial_{X_1}+\partial_{X_2})\sigma_z, 
\label{h2}\\
\mathcal{H}_3&=&-\frac{1}{2}\partial_{X_1}^2-\frac{1}{2}\partial_{X_2}^2 +\tilde{\mathcal{H}}_{\rm int}. 
\label{h3}
\end{eqnarray}
%
The 
interaction Hamiltonian $\tilde{\mathcal{H}}_{\rm int}$ is given by:
\begin{eqnarray}
\tilde{\mathcal{H}}_{\rm int}&=&\left [ \begin{array}{c c}
h_1+\beta h_2& 0 \\
0 & \beta h_1+h_2
\end{array}
\right], 
\label{Alphat}
\end{eqnarray}
where we have used 
$h_{i}=|u_{+}^{(1,i)}\phi^{(1)}_+|^2+|u_{-}^{(1,i)}\phi^{(1)}_-|^2 
+2{\rm Re}\{u_{+}^{(1,i)}u_{-}^{(1,i)}\phi^{(1)}_+\bar{\phi}_-^{(1)}\}$, and 
$u_{\pm}^{(1,i)}$ is the $i$-th component of the vector $\mathbf{u}_\pm^{(1)}$.
Inserting Eqs.~(\ref{expansion})-(\ref{h3})
into Eq.~(\ref{ham}), we obtain the following system of equations at the first three orders in $\epsilon$:
%
\begin{eqnarray}
\mathcal{O}(\epsilon):&\mathcal{L}(e^{ikx} \mathbf{u}_{\pm}^{(1)} \phi_{\pm}^{(1)})&=0, 
\label{ord1}\\
\mathcal{O}(\epsilon^2):&\mathcal{L}(e^{ikx}\mathbf{u}_{\pm}^{(2)} \phi_{\pm}^{(2)})&=\mathcal{H}_2(e^{ikx}\mathbf{u}_{\pm}^{(1)}\phi_{\pm}^{(1)}), 
\label{ord2}\\
\mathcal{O}(\epsilon^3):&\mathcal{L}(e^{ikx}\mathbf{u}_{\pm}^{(3)} \phi_{\pm}^{(3)})&=\mathcal{H}_2(e^{ikx} \mathbf{u}_{\pm}^{(2)} \phi_{\pm}^{(2)})
\nonumber \\ 
&&+(-i\partial_T + \mathcal{H}_3)(e^{ikx}\mathbf{u}_{\pm}^{(1)} \phi_{\pm}^{(1)}), \nonumber \\
\label{ord3}
\end{eqnarray}
where $\mathcal{L}=i\partial_t-\mathcal{H}_1$. 
Observing that the operator $\mathcal{L}$ does not act on the scalars $\phi_\pm^{(n)}$ (which depend only on the slow variables),
we find that Eq.~(\ref{ord1}) admits plane wave solutions of the form $\mathbf{u}^{(1)}_\pm =\mathbf{R}_\pm\exp(-i\omega_{\pm} t)$,  
where $\mathbf{R}_{\pm}$ are constant vectors satisfying the equation  
$\mathbf{W_\pm}\mathbf{R}_\pm=0$, where
\begin{eqnarray}
\mathbf{W_{\pm}}\!\!\!&=\!\!\!&	\left [ \begin{array}{c c}
\omega_\pm-k^2/2-kk_L-\delta & -\Omega \\
-\Omega & \omega_\pm-k^2/2+kk_L+\delta
\end{array}
\right]. \nonumber \\
\end{eqnarray}
Thus, the solvability condition of Eq.~(\ref{ord1}), ${\rm det}\mathbf{W_{\pm}}=0$, yields the dispersion relation of the linear problem:
\begin{eqnarray}
&\omega_{\pm}(k) = \frac{1}{2} k^2 \pm\sqrt{(k_Lk+\delta)^2+\Omega^2}.
\label{dr} 
\end{eqnarray}
Additionally we obtain the vectors $\mathbf{R}_{\pm}=c_\pm|\pm\rangle$, 
where $c_\pm$ are arbitrary constants, and $|\pm\rangle$ are the linear eigenvecrtors:
\begin{eqnarray}
|+\rangle=\left(\begin{array}{c}
\cos\theta\\
\sin\theta
\end{array}
\right), \quad
|-\rangle=\left(\begin{array}{c}
-\sin\theta\\
\cos\theta
\end{array}
\right). 
\label{eigen}
\end{eqnarray}
Here, the angle $\theta(k)=\cos^{-1}\left(1+Q^2\right)^{-1/2}$ sets the relative amplitude of each spin component, 
with the parameter $Q$ being given by $Q\equiv\Omega^{-1}[\sqrt{(k_Lk+\delta)^2+\Omega^2}-(k_Lk+\delta)]$.
%
 
Next we consider solutions of Eq.~(\ref{ord2}) in the form $\mathbf{u}_\pm^{(2)}=\mathbf{P}_\pm\exp(-i\omega_{\pm} t),$
leading to the following equation: 
\begin{eqnarray}
\mathbf{W_\pm}\mathbf{P}_\pm \phi_{\pm}^{(2)}&=&i\partial_{X_{1,2}}\phi_{\pm}^{(1)}\mathbf{W_\pm'}|\pm\rangle, \label{ord21} 
\end{eqnarray}
where primes denote differentiation with respect to the momentum $k$.
Projecting Eq.~(\ref{ord21}) to the nullspace of $\mathbf{W_\pm}$, 
we obtain the 
compatibility condition
$\langle\pm|\mathbf{W_\pm '}|\pm\rangle=0$, which yields the group velocities $v_g^{\pm} \equiv \omega_{\pm}'$ for the two energy bands:
\begin{eqnarray}
v_g^{\pm}=k\pm k_L[1-2\cos^2\theta(k)].
\label{grv}
\end{eqnarray}
The above equation indicates that, for any finite $k_L$, there exists a wavenumber $k_c=-\delta/k_L$ (for $\theta(k_c)=\pi/4$),
such that the two group velocities become equal $v_g^+=v_g^-\equiv v_g =k_c$. This result will be used below, 
to find solutions which travel together and exhibit persistent ZB oscillations. From Eqs.~(\ref{ord21}) we also obtain 
the solutions at $\mathcal{O}(\epsilon^2)$: 
\begin{eqnarray}
\mathbf{u}_\pm^{(2)} =\frac{\partial|\pm\rangle}{\partial k}\exp(-i\omega_{\pm} t), \quad
\phi_{\pm}^{(2)}=-i\partial_{X_{\pm}}\phi_{\pm}^{(1)}.
\end{eqnarray}  
Note that up to the order $\mathcal{O}(\epsilon^2)$, solutions in the two bands $\omega_{\pm}$ are decoupled 
as per the asymptotic expansion. However, at the order $\mathcal{O}(\epsilon^3)$, 
the compatibility condition of Eq.~(\ref{ord3}), 
\begin{eqnarray}
\langle\pm|\left[-i\partial_T-\frac{1}{2}\left(1+\mathbf{W}'_\pm\frac{\partial}{\partial k}\right)
\partial^2_{X_{\pm}}+\tilde{\mathcal{H}}_{\rm int}\right]\phi_{\pm}^{(1)}|\pm\rangle,
\label{comp2}
\end{eqnarray}
%
consists of two equations for $\phi_\pm^{(1)}$ which are nonlinearly coupled via the interaction Hamiltonian 
$\tilde{\mathcal{H}}_{\rm int}$. In the case of $k=k_c$, corresponding to equal group velocities $v_g^{+}=v_g^{-}$ at each energy band, 
these equations can be written as: 
\begin{eqnarray}
\!\!\!\!\!\!i\partial_T\phi_+&=&-\frac{\omega_+''}{2}\partial^2_X\phi_+ +\gamma(|c_+\phi_+|^2+|c_-\phi_-|^2)\phi_+,
\label{cnls11}\\
\!\!\!\!\!\!i\partial_T\phi_-&=&-\frac{\omega_-''}{2}\partial^2_X\phi_- +\gamma(|c_+\phi_+|^2+|c_-\phi_-|^2)\phi_-,	
\label{cnls22}
\end{eqnarray}
where superscripts have been dropped for simplicity; here, $\gamma=(1+\beta)/2>0$ is the common coupling constant, 
and $X_+=X_- \equiv X=\epsilon(x-v_g t)$. 

\section{Beating dark-dark solitons}
The system of coupled equations~(\ref{cnls11})-(\ref{cnls22}) is the main result of this work, and below we will
briefly comment on some of its properties. These equations describe the evolution of two wavefunctions, each residing in a 
different energy band $\omega_{\pm}$, and interacting through the nonlinear terms; note that the condensate components
consist of linear combinations of both $\phi_+$ and $\phi_-$. 
In the limit of $c_+c_-=0$, when only one energy band is populated, 
Eqs.~(\ref{cnls11})-(\ref{cnls22}) reduce to two decoupled NLS equations for $\phi_+$ or $\phi_-$ respectively. In this limiting
case, localized nonlinear solutions of these NLS equations take the form of solitons, residing in one of the 
two energy bands $\omega_\pm$~\cite{prlva,eplva}. In the more general case of $c_+ c_- \ne 0$, localized solutions of
Eqs.~(\ref{cnls11})-(\ref{cnls22}) provide vector solitons for the original equations, with each soliton component $\phi_\pm$,
residing in the respective energy band, $\omega_\pm$. 
Below, without loss of generality we fix $c_\pm=1$.
%
%

When the dispersion coefficients are equal, $\omega''_+=\omega''_-$, the system reduces to the completely integrable
Manakov model~\cite{manakov}, which possesses exact analytical $N$-soliton solutions~\cite{ohta}.
In the more general case the system of  Eqs.~(\ref{cnls11})-(\ref{cnls22}) is characterized by two distinct length scales, stemming from the
terms $\omega_+''$ and $\omega_-''$ which in turn depend on the parameters of the original problem $k_L$ and $\Omega$.
In our case, however, 
it turns out that 
$\omega''_+(k_c)\ne\omega''_-(k_c)$, where $\omega''_+(k_c)>0$ and $\omega''_-(k_c)$ is positive (negative) for $\Omega>k_L^2$ 
($\Omega<k_L^2$); thus, generally, exact analytical soliton solutions do not exist. 
Below we focus on the case 
$\omega''_+(k_c)\omega''_-(k_c)>0$, and obtain numerically 
stationary solutions of Eqs.~(\ref{cnls11})-(\ref{cnls22}), 
of the form $\phi_{\pm}(X,T)=\Phi_{\pm}(X)\exp(i\mu_\pm T)$,  
where $\mu_\pm$ are effective chemical potentials. 


\begin{figure}[tbp]
\includegraphics[width=8.5cm]{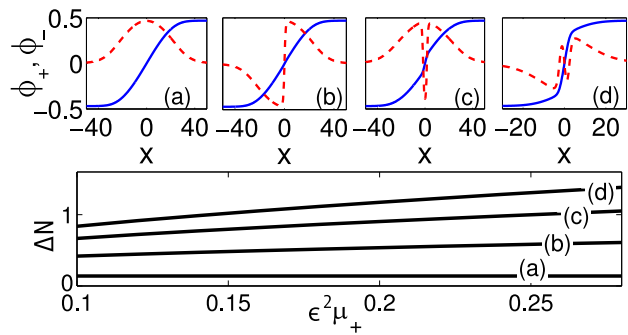}
\caption{(Color online) Top panel: profiles of soliton solutions of Eqs.~(\ref{cnls11})-(\ref{cnls22}), for $\epsilon^2\mu_+=0.26$;
solid (blue) and dashed (red) lines correspond to $\Phi_+$ and $\Phi_-$ respectively.
Bottom panel: the normalized difference  $\Delta N =\int[(\mu_+-|\Phi_+|^2)-|\Phi_-|^2]dx/\int(\mu_+-|\Phi_+|^2)dx$ 
between the number of atoms of dark and bright components, as a function of $\mu_+$ for the soliton branches (a)-(d). 
Parameter values are: $\Omega/k_L^2=1.25$, $\delta=0$, $k=0$, and $\epsilon=0.1$.}
\label{branch}
\end{figure}

Our first initial guess --motivated by the Manakov limit-- 
is:
$$\Phi_+(X) \propto \tanh(b_+ X), \quad 
\Phi_-(X) \propto {\rm sech}(b_+ X),$$ 
corresponding to 
a dark-bright soliton of width $b_+$. Using this initial guess in a fixed point iteration,
we numerically obtain a soliton branch, whose 
spatial profile is shown in Fig.~\ref{branch}(a)  
as a function of the original spatial variable $x$ (this corresponds to branch (a) in the bottom panel 
of the same figure).

Motivated by the existence of two length scales of the system, 
for the same form of $\Phi_+(X)$, we look for other solutions assuming
that:
$$\Phi_-(X) \propto {\rm sech}(b_+ X)\tanh(b_- X),$$ 
where $b_->b_+$ is the width of a dark soliton ``imprinted'' in the 
bright component. Such so-called  ``twisted'' (odd parity) solitons have been previously predicted 
to arise e.g., in single-component trapped BECs~\cite{NJP,FRM} and in nonlinear 
fiber optics~\cite{pare} settings.
This way, we are able to obtain solutions belonging to branch (b), an example of which  
is shown in Fig.~\ref{branch}(b).
Higher-order excited states were also obtained using the same pattern. An initial guess of the form: 
$$\Phi_-(X) \propto {\rm sech}(b_+ X)\tanh[b_- (X+X_0)] \tanh[b_- (X-X_0)],$$ 
leads to solitons of branch (c) shown in Fig.~\ref{branch}. Finally,  
using: 
\begin{eqnarray}
\Phi_-(X) &\propto& {\rm sech}(b_+ X) \tanh[b_- (X+X_0)] 
\nonumber \\
&\times& \tanh(b_- X) \tanh[b_- (X-X_0)], \nonumber
\end{eqnarray}
we obtain branch (d) of Fig.~\ref{branch}. 
Here, $X_0$ denotes the displacements of the dark solitons from the origin, and these solutions correspond to two and three dark solitons
``imprinted'' on the  bright soliton component. 
For the examples of Fig.~\ref{branch}, we used  
$\Omega/k_L^2=1.25$ and $\delta=0$ (corresponding to $\omega_+''=1.8$ and  
$\omega_-''=0.2$), as well as $\epsilon^2\mu_+=0.2$ and $\epsilon^2\mu_-=0.19$; 
nevertheless, branches of relevant states were 
also found for different $\mu_\pm$ [cf. bottom panel of Fig.~\ref{branch}]. 

We have also confirmed the existence of these states for $\delta \ne 0$ corresponding to $v_g\ne 0$ (see below). 
Notice that other soliton solutions, corresponding to different initial choices (e.g., multiple dark-bright solitons \cite{DD}) 
and/or different ratios $\omega_+''/\omega_-''$ (controlling the number of imprinted dark solitons in the bright component)
and importantly even for different signs of this quantity, can be found as well.
\begin{figure}[tbp]
\includegraphics[width=8.5cm]{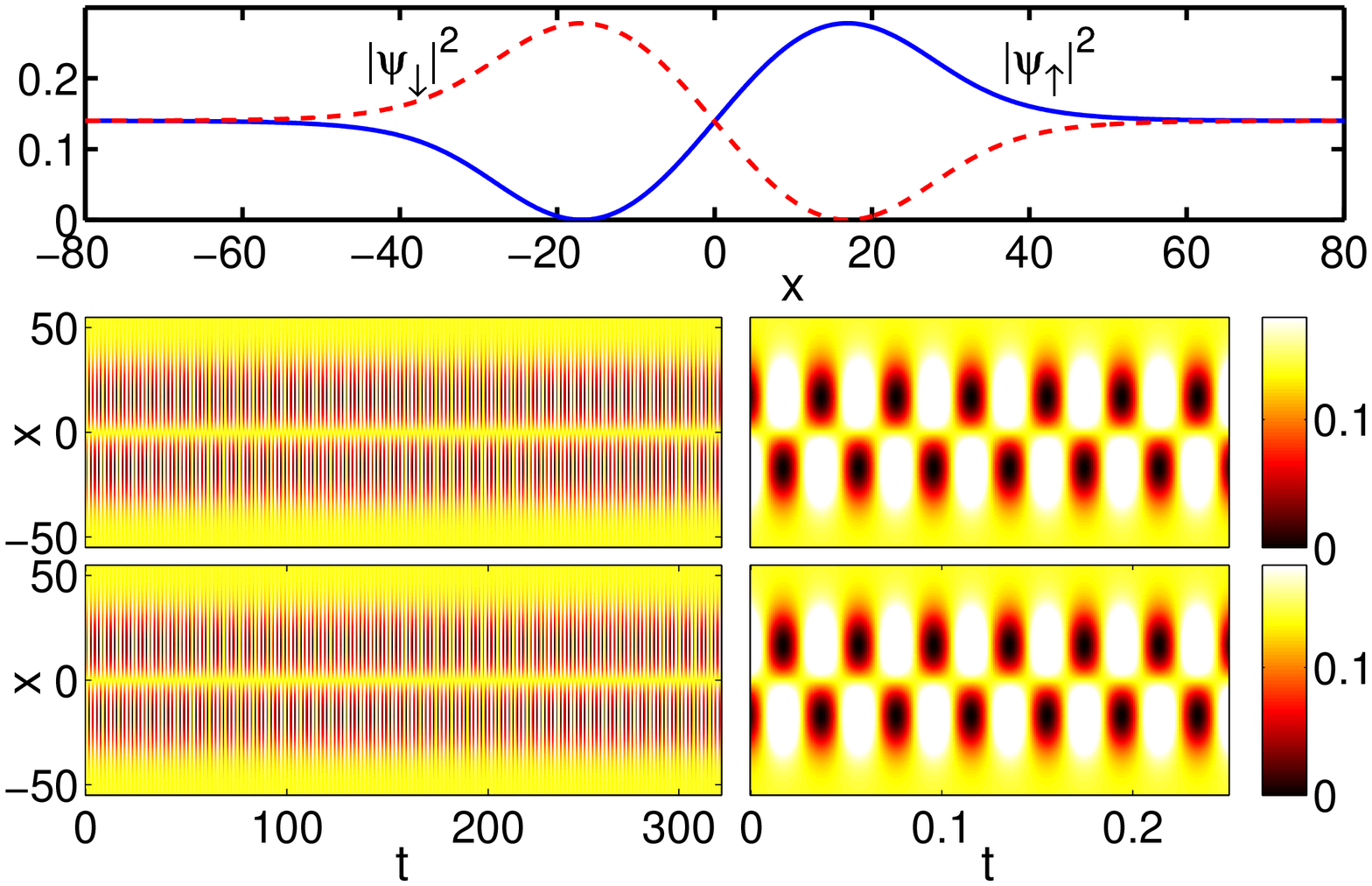}
\includegraphics[width=8.5cm]{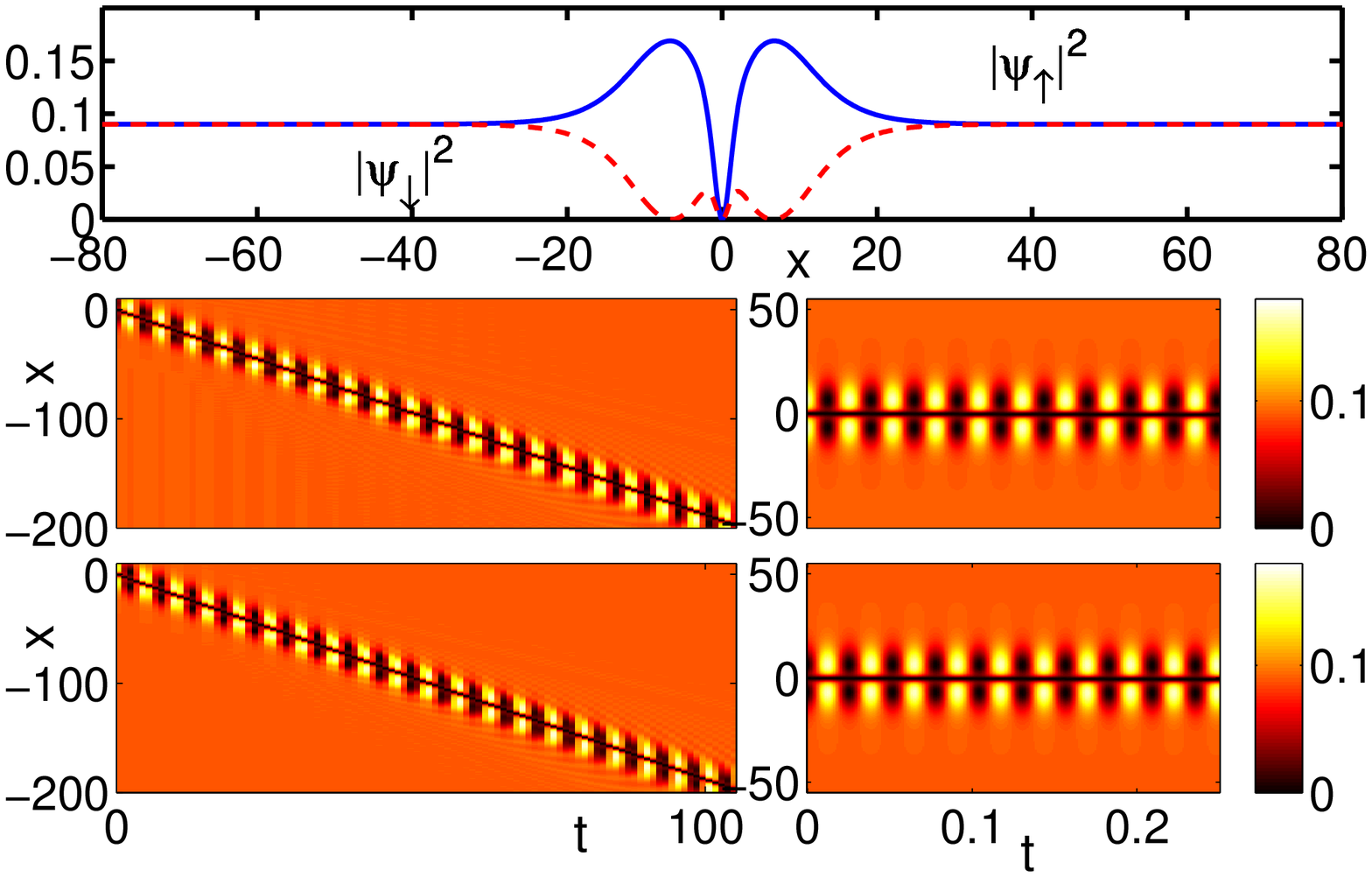}
\caption{(Color online) Top row: initial condition corresponding to a soliton of branch~(a)  
of Fig.~\ref{branch}. Second and third row: evolution of the densities 
$|\Psi_\uparrow|^2$ and $|\Psi_\downarrow|^2$ respectively (left), and a close-up showing in detail  
the oscillations (right). Parameters used are the same as in Fig.~\ref{branch} with $\epsilon^2\mu_+=0.1$, $\epsilon^2\mu_-=0.095$. 
The lower three rows are similar to the above, but for a soliton of branch~(b) with $k=-2$, and other parameters 
$\Omega/k_L^2=1.8$, $\epsilon^2\mu_+=0.2$, and $\epsilon^2\mu_-=0.195$.}
\label{db0}
\end{figure}

Having derived soliton solutions of Eqs.~(\ref{cnls11})-(\ref{cnls22}), we can employ Eq.~(\ref{expansion}) and construct 
localized nonlinear solutions of Eq.~(\ref{ham}).
Such solutions, in the form of solitons occupying both energy bands are given by:
%
\begin{eqnarray}
\!\!\!
|\psi_{\uparrow, \downarrow}|^2 \approx \epsilon^2 \left[\frac{1}{2} \rho \mp {\rm Re} \{\phi_+\bar{\phi}_-\}
\cos[(\delta\omega -\epsilon^2\delta\mu)t]\right], 
\label{evolpsi}
\end{eqnarray}
where $\rho \equiv |\phi_+|^2+|\phi_-|^2$ and	 $\delta \mu =\mu_+-\mu_-$.
The soliton densities
are time-periodic, with a frequency 
$\omega_{\rm ZB} \equiv \delta\omega -\epsilon^2\delta\mu$, where $\delta\omega \equiv \omega_+ -\omega_-$ is the usual 
ZB frequency \cite{zbbec2}. In our case, this frequency is slightly modified due to nonlinearity, and 
will be called, hereafter, ``nonlinear ZB frequency''. Notice that these solitons are termed 
``beating'' since their densities oscillate in time 
around the soliton core --but with the asymptotics at $x\rightarrow \pm \infty$ being constant.


The dynamics of the obtained solitons is studied by means of numerical integration of 
Eq.~(\ref{ham}), using as initial conditions the solitons of Eqs.~(\ref{evolpsi}), with $\phi_\pm$ corresponding to 
branches (a) and (b). In Fig.~\ref{db0} the first three rows, corresponding to a soliton belonging to branch (a), 
depict the initial density profile (top panel), and contour plots showing the evolution of individual 
densities $|\Psi_\uparrow|^2$ and $|\Psi_\downarrow|^2$ (middle and bottom panels, respectively). 
This state
exhibits density oscillations (cf. middle and bottom right panels) with a frequency $\omega_{\rm ZB}$, 
such that the individual numbers of atoms 
$N_{\downarrow,\uparrow}$
are constant. This can be quantified by using the mean value
of the $z$-spin component, $\langle \sigma_z\rangle=(N_\downarrow-N_\uparrow)/(N_\downarrow+N_\uparrow)$. 
Employing Eq.~(\ref{evolpsi}), we find that the latter is given by: 
\begin{eqnarray}
\!\!\!\!
\langle\sigma_z(t)\rangle=A \cos(\omega_{\rm ZB}t), \quad \!\!\!\!
A=\frac{\int{\rm Re} \{\phi_+(x)\bar{\phi}_-(x)\} dx}{\int \rho(x)dx}.
\label{spinz2}
\end{eqnarray}
It is now clear that for solitons of branch~(a), with an odd parity of the product $\phi_+\phi_-$, one gets $A=0$ and, thus, 
a zero mean value of $\langle \sigma_z\rangle$.

Here we should note that although solitons of branch~(a) are similar to 
beating dark-dark solitons observed experimentally in~\cite{engels3,DD}, there exists a substantial difference: 
the breathing behavior of all states presented herein is characterized by an oscillation frequency 
equal to the nonlinear ZB frequency, corresponding to solitons in different bands.

The bottom three rows of Fig.~\ref{db0} show the evolution of a soliton belonging to 
branch (b), with a finite momentum ($k=-2$). This state, having a density dip located at the origin,
exhibits different dynamical behavior: the number of atoms in each component now oscillates,
since $\phi_+\phi_-$ is even. Thus, $\langle \sigma_z\rangle$ is now nonzero (since $A\ne 0$), and 
oscillates with the nonlinear ZB frequency $\omega_{\rm ZB}$.
Summarizing the above results we find that: solitons corresponding to branches with odd parity of the product $\phi_+\phi_-$
have vanishing z-spin component (unpolarized) and branches with even parity 
thereof perform spin oscillations, 
with the frequency  $\omega_{\rm ZB}$. We stress that solitons of all branches were found to be persistent in our 
numerical simulations, at least up to a normalized time $t \approx 500$.

\section{ZB oscillations}

After considering the dynamics of the obtained soliton states, we will now 
provide a  connection between spin oscillations and  ZB oscillations of the different branches.
The center of mass velocity can be obtained
%
by using the Heisenberg equation of motion, and considering the space operator $\hat{x}$ as time dependent \cite{zbbec2}: 
\begin{eqnarray}
\langle v(t)\rangle=\langle \frac{d\hat{x}(t)}{dt} \rangle = \langle i\left[\hat{x},\mathcal{H}_1\right] \rangle=k+k_L\langle\sigma_z\rangle,
\label{meanv2}
\end{eqnarray}
where $[\hat{x},\mathcal{H}_1] \equiv \hat{x}\mathcal{H}_1-\mathcal{H}_1 \hat{x}$ is the usual commutator, 
and  we have used the fact that the mean
value of the momentum $\langle p \rangle\approx k$ for 
the soliton branches obtained herein [note that in the fast scales our solutions have the form of plane waves
$\sim\exp(ikx)$].  
%

It is clear from Eq.~(\ref{meanv2}) that any spin oscillations $\langle \sigma_z(t)\rangle$
(i.e., exchange of atoms between components) will result in a time-dependent velocity, which is the signature of the ZB phenomenon 
--i.e., the velocity oscillates due to spin oscillations.
According to the above analysis for $\langle \sigma_z(t)\rangle$ (where it was shown that 
spin oscillations may occur, depending on the parity of the product $\phi_+\phi_-$), we can now end up with the 
following conclusion: 
%
%
%
solitons of branches (b) and (d) 
exhibit ZB oscillations, while solitons of branches (a) and (c) do not. 
Note that if $k_L=0$ then Rabi oscillations occur due to the linear coupling between components, but 
the mean velocity does not oscillate; this distinguishes Rabi from ZB oscillations.

\begin{figure}[tbp]
\includegraphics[width=8.4cm]{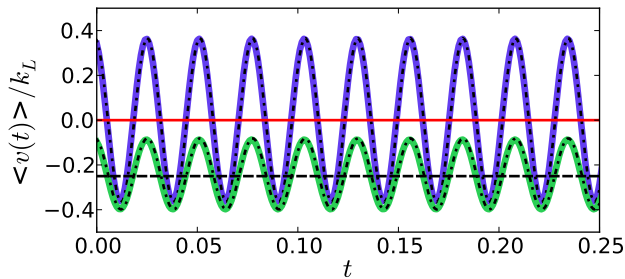}
\caption{(Color online) The time dependence of the mean velocity $\langle v(t) \rangle$ 
normalized to the wavenumber $k_L$, for different soliton solutions. Solid (red) line at $\langle v(t) \rangle=0$ 
corresponds to a stationary soliton of branch (a).
The larger amplitude solid (blue) and smaller amplitude solid (green) lines correspond 
to solitons of  branch (b) at $k=0$ and $k=-2$, respectively. The dotted (black) lines correspond to the semi-analytical result of Eq.~(\ref{meanv2}), and the dashed (black) line
indicates the shift from the center to the soliton with $k=-2$.
Parameters used are $\Omega/k_L^2=1.8$ and $\epsilon^2\mu_+=0.26$.
}
\label{spin}
\end{figure}

The analytical prediction of Eqs.~(\ref{spinz2}) and (\ref{meanv2}) was tested against 
numerical simulations for different soliton states. 
Particularly, in Fig.~\ref{spin}, we show the time dependence of $\langle v(t)\rangle$ for   
solitons of branches (a) and (b), and 
%
%
%
depict results obtained numerically (solid lines) and semi-analytically (dotted lines):  
for the former, $\langle v(t)\rangle$ was found as the time derivative of the center of mass coordinate, and for the latter 
Eqs.~(\ref{spinz2})-(\ref{meanv2}) were used. The solid (red) curve at $\langle v(t)\rangle=0$ corresponds to a soliton of
branch (a), while the larger (smaller) amplitude curve corresponds to a stationary (moving) soliton of branch (b)  
with $k=\delta=0$ ($k=-2$ and $\delta=16$); in all cases $\Omega/k_L^2=1.8$. Note that the velocity oscillations 
for the soliton with $v_g\ne 0$, are performed around the value $k/k_L=-0.25$ [cf. Eq.~(\ref{meanv2})], depicted 
by the dashed (black) horizontal line. Clearly, 
the analytical predictions for the occurrence and the frequency, of ZB oscillations are confirmed, while there is 
a good agreement between numerical and semi-analytical results for $\langle v(t)\rangle$. 

Notice that although 
Eq.~(\ref{meanv2}) stems from the linear theory, it can capture accurately the mean velocity of 
solitons. This should not be surprising, as our analytical approximations refer to a weakly nonlinear theory 
(in the sense that solitons are of small-amplitude).
%

%
%
%

\section{Discussion and conclusions} 

In this work, we illustrated a prototypical example of
nonlinear coherent states, in the form of beating dark-dark (DD) solitons,
occupying both energy bands of the spectrum of a spin-orbit coupled (SOC) BEC,
and discussed their connection with Zitterbewegung oscillations. 
Using a multiscale expansion method, we obtained
two coupled NLS equations describing the evolution of 
nonlinear wavepackets in the two different energy bands of a SOC-BEC. 
Solutions of this system were found, and were intriguing in their own
right, featuring {\it embedded} families of bright, twisted or higher
excited solitons inside a dark soliton. These unusual families of
solitons were made possible by the disparity in dispersion between
the different bands (at the equal group velocity point) and 
accordingly featured two distinct length scales. 

These solutions were then used to obtain different soliton branches of the initial system,
with densities oscillating with the characteristic ZB frequency. Numerical results,
provided the proof-of-principle of that the derived families DD-soliton families, where appropriate,
are robust and long-lived.
Each branch was then characterized with respect to its spin dynamics, and it was shown that 
odd branches of the product $\phi_+ \phi_-$ have zero z-spin 
component while even branches of this product exhibit spin oscillations.
We have shown that branches with spin oscillations, exhibit also ZB oscillations
due to spin-orbit coupling. Our analytical predictions for the dynamics of the ZB oscillations 
were also confirmed numerically. 

The derived dark solitons should, in principle, be observable 
in experiments with spin-orbit coupled BECs, similar to the ones
already performed in~\cite{zbbec2}, and pose an exciting challenge
for the current experimental state-of-the-art. 
The generality of the considered model and of our methodology, offer a deeper understanding 
of solitons dynamics in models with Dirac-like energy band (or other multi-band) structure: these include 
the massive Thirring model~\cite{thirring}, models in 
nonlinear optics describing optical fiber gratings~\cite{christo}, birefringent optical fibers
and coupled optical waveguides~\cite{boris,longhi}, or even acoustic 
surface waves~\cite{acoustics}.

There exists a number of possible future research directions that are suggested by this study. 
These include, for instance, the identification of novel nonlinear structures in multi-band systems and 
the study of their ZB oscillations. Such structures could be composed by other soliton states, 
e.g., bright solitons, as well as vortices, vortex clusters, or vortex-bright solitons \cite{panos} 
in higher-dimensional systems.

\section*{ACKNOWLEDGEMENTS} The work of D.J.F. was partially supported by the Special Account for 
Research Grants of the University of Athens.
PGK acknowledges support from the Alexander von
Humboldt Foundation, the Binational Science Foundation 
under grant 2010239, NSF-CMMI-1000337, 
NSF-DMS-1312856, FP7, Marie Curie Actions, People, International
Research Staff Exchange Scheme (IRSES-606096) and from the US-AFOSR under grant
FA9550-12-10332.

\end{document}